# Accelerating Copolymer Inverse Design using AI Gaming algorithm

Tarak K Patra[1*], Troy D. Loeffler[2] and Subramanian K R S Sankaranarayanan[2, 3*]

[1]Department of Chemical Engineering, Indian Institute of Technology Madras, Chennai, Tamil Nadu 600036, India
[2]Center for Nanoscale Materials, Argonne National Laboratory, Lemont, Illinois 60439, USA
[3]Department of Mechanical and Industrial Engineering, University of Illinois at Chicago, Chicago, Illinois 60607, USA

**Abstract**

There exists a broad class of sequencing problems, for example, in proteins and polymers that can be formulated as a heuristic search algorithm that involve decision making akin to a computer game. AI gaming algorithms such as Monte Carlo tree search (MCTS) gained prominence after their exemplary performance in the computer Go game and are decision trees aimed at identifying the path (moves) that should be taken by the policy to reach the final winning or optimal solution. Major challenges in inverse sequencing problems are that the materials search space is extremely vast and property evaluation for each sequence is computationally demanding. Reaching an optimal solution by minimizing the total number of evaluations in a given design cycle is therefore highly desirable. We demonstrate that one can adopt this approach for solving the sequencing problem by developing and growing a decision tree, where each node in the tree is a candidate sequence whose fitness is directly evaluated by molecular simulations. We interface MCTS with MD simulations and use a representative example of designing a copolymer compatibilizer, where the goal is to identify sequence specific copolymers that lead to zero interfacial energy between two immiscible homopolymers. We apply the MCTS algorithm to polymer chain lengths varying from 10-mer to 30-mer, wherein the overall search space varies from $2^{10}$ (1024) to $2^{30}$ (~1 billion). In each case, we identify a target sequence that leads to zero interfacial energy within a few hundred evaluations demonstrating the scalability and efficiency of MCTS in exploring practical materials design problems with exceedingly vast chemical/material search space. Our MCTS-MD framework can be easily extended to several other polymer and protein inverse design problems, in particular, for cases where sequence-property data is either unavailable and/or is resource intensive.

*E-mail: tpatra@iitm.ac.in (TKP), skrssank@uic.edu (SKRSS)

**Introduction**

There exists a broad class of soft-materials such as proteins and polymers where the arrangement of moieties *i.e.* the sequence plays a critical role in determining their functionality. For instance, activities and functionalities of DNA and other biomolecules are determined by the exact sequence of amino acids and other chemical moieties in their back bones.[24–26] As an example, the arrangement of amino acid sequence in viruses plays a key role in determining their mutations and hence the effectiveness of the drugs or vaccine used to treat them. Likewise, several recent studies indicate that the sequence specificity of the constituents chemical moieties of a copolymer can lead to more efficient materials – their thermodynamic properties such as miscibility and surface tension as well as structure/morphology is strongly influenced by the sequence in oligomers. It is therefore not surprising that a lot of effort has focused on controlling the sequence specificity in polymers, proteins and other biomolecules.

On the experimental front, progress in synthetic chemistry has enabled us to exercise an unprecedented control over sequences in copolymers - such precision polymers remain an area of major focus in current fundamental and applied polymer research.[27–29] Copolymers are a special class of polymers that comprise of more than one type of chemical species; and shows a reach phase behaviour[15–17] and tunability in its thermophysical properties.[18–23] These copolymers are usually characterized by their mean block length and mass fraction. One of the area where sequence specificity is found to play an import role is the use of copolymer as interfacial compatibilizers.[2,30] Copolymer compatibilizers are commonly employed to improve the thermodynamic stability of polymer interfaces, and they therefore have wide applicability in emulsions and composite materials.[19,31]

A major challenge in the design of sequence specific polymers lies in the vast combinatorial space that precludes efficient exploration. For instance, a polymer chain with $n$ number of possible monomers and $m$ type of monomers will have $n^m$ possible combinations that can be likely explored. Even for a binary polymer i.e. 2 types with a chain length of ~30 units, the total combinations possible are $2^{30}$ which is close to 1 billion. Given such an enormous sequence space that needs to be explored, it is highly desirable to minimize the number of trials needed to arrive at a sequence that corresponds to a desired target property. Fortunately, the emergence of artificial intelligence (AI) positions us uniquely to solve this seemingly intractable inverse design problem.

AI and machine learning (ML) can be combined with molecular simulations to solve the inverse problem and accelerate materials discovery/design. Molecular simulations such as molecular dynamics (MD) are powerful techniques to evaluate the sequence-property relationships. Typically, MD simulations can sample the configurational & property space and create adequately large structure-property training datasets of a materials. On the other hand, ML methods can very efficiently screen this extensive dataset and identify sequences or configurations that correspond to desired optimal material properties. Such inverse problems have been traditionally addressed using evolutionary methods such as genetic algorithms (GA) or Bayesian optimization (BO) – these optimizers are combined with MD simulations to identify target properties for a wide range of materials from inorganic to semiconductor to polymers.[1–6] In the last few years, such combination of MD and ML have been successfully deployed to explore the vast configurational space of materials. Their widespread application to problems of practical interest, however, requires addressing two bottlenecks. First, the GA and BO methods exhibit poor scalability as the design space increases. Typically, the search space of most practical sequencing problems exceeds several millions and higher. Second, they find difficulties in surmounting suboptimal solutions and tends to slow down near the optimal points. Third, each property evaluation in many design problems is computationally intensive (trajectories over several tens of nanoseconds and more), which precludes high-throughput exploration. Within a typical MD-ML materials design framework, one often requires several thousands to millions of direct evaluations or computations of materials properties. In the context of soft materials, this poses a major limitation, for instance, when the MD calculations for each sequence are computational very expensive such as, for instance, computing the thermophysical properties of polymeric materials. We note that the relaxation in polymeric materials is inherently slow and requires significantly long MD simulations to calculate their equilibrium properties. A key challenge in accelerating computer aided molecular-scale polymer design and address the sequence problems in materials design is to significantly reduce the number of direct computational evaluation of materials property that are required to identify an optimal candidate corresponding to the target property.

The advent of big data analytics and powerful supercomputers have brought AI and ML techniques that can address the above challenges in materials design. In this front, Monte Carlo tress search (MCTS) has emerged as a powerful global optimization method that has found wide-spread applications in computer games such as Alpha Go, games such as Bridge, Poker and many other video games.[7,8] MCTS is a probabilistic and heuristic search algorithm that integrates a tree search algorithm with machine learning principles of reinforcement learning. MCTS is a decision tree-based approach that builds a shallow tree of nodes where each node represents a point in the search space and downstream pathways are generated by a rollout procedure. The algorithm simultaneously explores potentially better pathways to reach the optimal point in a search space and exploits a single pathway that has the greatest estimate value of the search function. This combination of exploration *vs.* exploitation and an appropriate trade-off mechanism between them are found to be the most efficient strategy of identifying optimal point for a given function. An advantage of the MCTS is that if the search gets trapped in a metastable or suboptimal point, it can quickly find another pathway by growing other branches of the tree utilizing the trade-off mechanism between exploration and exploitation. MCTS has been successfully adopted for material science problems such as predicting silicon-germanium alloy structure with optimal thermal conductance,[9,10] and discovering new synthetic routes of making organic molecules,[11] optimal atoms segregation at grain boundary,[12] predicting organic molecules with optimal partition coefficient and other properties,[13] and enhancing biomolecular sampling.[14]

Here, we draw inspiration from the success of AI algorithms such as MCTS in computer games and aim to develop a design algorithm for sequence problems that are fast (time to solution) as well as highly scalable. We focus on a representative albeit complex polymer inverse design problem, viz., design of sequence of copolymer molecules that corresponds to a user-desired property. Our goal is to design the sequence of the compatibilizer that minimizes the interfacial tension between immiscible polymers. Block copolymers and random copolymers have long been used as compatibilizers that reduce interfacial tension between immiscible polymers and improve the stability of the composite materials.[30,32] Recently, evolutionary search based on MD simulations (MD-GA) have identified sequence specific co-polymers that outperform block and random copolymers.[2] However, an optimal solution (a sequence specific copolymer) for a 20 bead polymer chain via evolutionary search required several thousands of MD simulations within the MD-GA framework. In practice, the polymer chains can often involve several tens of monomers to several hundreds necessitating algorithms that are efficient and scalable. In our design workflow, we interface MCTS with MD simulations that is used for evaluating the objective function for any specific sequence. Our MD simulation based Monte Carlo tree search (MD-MCTS) workflow rapidly identifies optimal sequence of copolymers corresponding to our desired interfacial tension. We demonstrate the scalability of our workflow by simulating chain lengths from 10 to 30 monomers – for each case the search required only a few hundred evaluations despite the search space extending from 512 to 0.5 billion, respectively. Our work demonstrates the success of AI in efficient and faster materials search and is applicable for a broad class of sequence related materials design problems.

**Model and Methodology**

**Molecular Dynamics of Polymers**: We use a generic coarse-grained model to represent two homopolymers A and B that are immiscible. The compatibilizer is a copolymer consist with both the A and B type moieties. Within this model system, two adjacent coarse-grained monomers of a polymer is connected by the Finitely Extensible Nonlinear Elastic (FENE) potential of the form

$$E = -\frac{1}{2}KR_0^2\left[1 - \left(\frac{r}{R_0}\right)^2\right].$$

Here, $k = 30\epsilon/\sigma^2$ and $R_0 = 1.5\sigma$. Any two monomers in the system is interacted via the Lennard-Jones (LJ) potential of the form

$$V(r_{ij}) = 4\epsilon_{ij}\left[\left(\frac{\sigma}{r_{ij}}\right)^{12} - \left(\frac{\sigma}{r_{ij}}\right)^6\right].$$

The $\epsilon_{ij}$ is the interaction energy between any two monomers $i$ and $j$. The size of all the monomers are σ. The LJ interaction is truncated at a cut-off distance $r_c = 2.5\sigma$ to represent attractive interaction among the monomers of homopolymers viz., A-A and B-B interactions. The immiscibility between homopolymer A and B is modelled by pure repulsion between A and B moieties. This is achieved by choosing $r_c = 2^{1/6}\sigma$ for A-B interaction. The orthogonal simulations box consist of a total of 693 homopolymer A and 693 homopolymer B that form two interfaces as shown in Figure 1a. The compatibilizer chains are placed at both the interfaces. The interfacial area between the two homopolymers ($36\sigma \times 36\sigma$) is kept constant during the simulations. Also the compatibilizer concentration at each interface which is defined as the compatibilizer monomers per unit area of an interface, is kept constant. Three case studies are conducted each for varying compatibilizer chain length. A total of 414, 207 and 138 compatibilizer copolymer chains of length N=10, 20 and 30, respectively, are placed at both the interfaces. This lead to a compatibilizer density of $1.59/\sigma^2$. All the systems consist of 36000 CG beads. The system is periodic in all three direction. All the simulations employ the Verlet time integration scheme[33] with a time step of $0.005\tau$. The Noose Hover thermostat and barostat[34] are employed to keep the temperature and pressure constant during the simulations.

**Property Evaluation**: All the MD calculations are conducted at a reduce temperature T=1 and zero pressure in the direction normal to the interface. During a MD summation, a system is initially equilibrated for $2\times10^6$ MD steps, followed by a production run of another $2\times10^6$ steps.

During the production run, pressure tensor data are collected, and the surface tension of a system is calculated as

$$\gamma_{12} = \langle \frac{L_z}{2}\left(P_{zz} - \frac{1}{2}(P_{xx} + P_{yy})\right)\rangle.$$

Here z is the direction normal to the interface, the in-equilibrium box length along z is represented by $L_z$. The Pxx, Pyy and Pzz are the pressure components along three directions. All the MD simulations are conducted using LAMMPS molecular dynamics simulation package.[35]

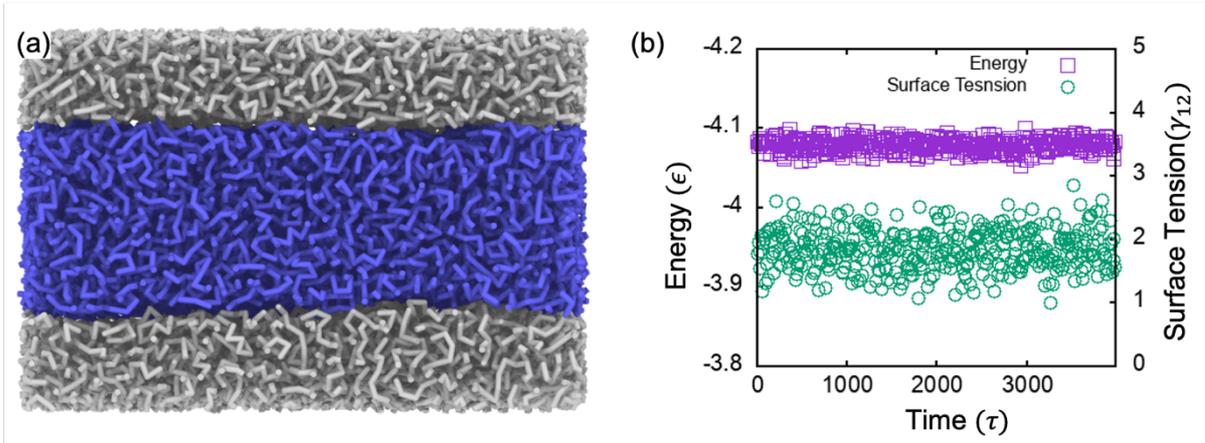

*Figure 1: A polymer blend: (a) MD snapshot of two immiscible homopolymers forming interfaces in a MD simulation box. (b) energy and surface tension of the system is shown as a function of time during the production run. The energy, time and surface tension values are in LJ units.*

**Material Systems**: We begin by modelling the base material where two immiscible homopolymers of type A and B form interfaces as shown by the MD snapshot in Figure 1a. The in-equilibrium energy and surface tension is shown as a function of time in Figure 1b. It suggests the system has reached a steady state within the equilibrium run and all the properties are calculated by time averaging of data collected in these equilibrium region of the trajectory. The time average energy and surface tension of the system are -4.09ε and 1.8ε/σ², respectively. Here, ε and σ are the unit of energy and length, respectively. Now we place copolymer chains which consist of both A and B type moieties at the interfaces of the simulation box. We explore the sequence of A and B type moieties in a given copolymer or compatibilizer that reduces the surface tension of the system. There are three compatibilizer chains are considered, they are of chain length N = 10, 20 and 30 as shown in Figure 2. The total number of possible candidate structures out of a binary chain is $C = 2^N/2$. We note that the denominator is to avoid the double counting of configurations as a sequence and its reverse sequence are identical in this context. Therefore, the total number of candidate structures or sequence in the search space are 512, 524288 and 536870912 for N= 10, 20 and 30, respectively. We seek to identify the sequence of moieties A and B that lead to lowest surface tension of the system for all the three cases by combining MD and MCTS.

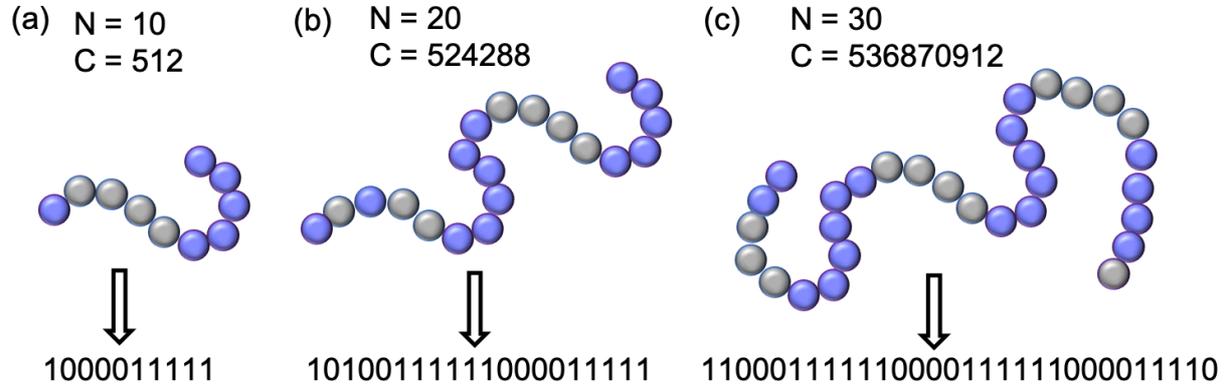

Figure 3: Binary Mapping of copolymers. Copolymer of length N=10,20 and 30 are shown in (a), (b) and (c), respectively. The grey and blue beads represent monomer of type A and B, respectively. The arrows point to the one dimensional binary strings where 0 and 1 correspond to monomer A and B, respectively. The binary string length is same as the copolymer chain length N. Here, the C represent total number of sequences possible for a given copolymer of size N.

**Monte Carlo Tree Search For Co-polymer Design:** The MD-MCTS workflow for exploring this search space is shown schematically in Fig. 3. Thus, the surface tension of the system in presence of compatibilizer chains can be written as $\gamma_{12} = \gamma_{12}(x)$. Here, $x \in \{0, 1\}^N$ represent a sequence of 0 and 1 of size N; and 0 and 1 correspond to moieties A and B, respectively. We conduct three MD-MCTS calculations each for a specific value of N. For a given N, the MD-MCTS begins with randomly generating a sequence of 0 and 1 of size N. This candidate serve as the root node of the search tree. The search tree is built in an incremental and iterative way as shown in Figure 3 until a pre-defined termination criteria is reached. Once the termination criteria is reached, the search is stopped and the best performing candidate is returned. Our termination criteria is set to be $\gamma_{12} \approx 0$. In each iteration, four steps – selection, expansion, simulations and back-propagation are carried out. A child node is selected during the selection process based on the upper confidence bound (UCB) score.[36] The UCB of a node is defined as $ucb_i = \frac{z_i}{v_i} + C\sqrt{\frac{2 \ln v_p}{v_i}}$. Here, $z_i$ is the accumulated merit of the node (i.e., the sum of the immediate merits of all the downstream nodes), and $v_i$ is the visit count of the node, $v_p$ is the visit count of the parent node and $C$ is a constant for balancing exploration and exploitation. The value of C is controlled adaptively at each node as $C = \frac{\sqrt{2}J}{4}(z_{max} - z_{min})$. Here, J is the meta parameter which is set to be one and it increases whenever the algorithm reach a "dead end" node to allow more exploration. At a dead-end node, the number of possible structures narrows to one. This happens when the numbers of k − 1 candidate structures reach the limit. Here, the J is updated as $J \leftarrow J + max\left\{\frac{T-t}{T}, 0.1\right\}$, where T is the total number of candidates to be evaluated, and $t$ is the number of candidates for which the surface tension is already evaluated. Whenever a new node is added, the variables are initialized as $v_i = w_i = f_i = 0$ and $u_i = \infty$. Next, we perform the expansion of the tree by adding child nodes to the selected node. In the simulation step, a playout is performed from each of the added children. We roll out 10 structures randomly during a playout from a child node. All the 10 structures are evaluated via MD simulations. Finally, in the back propagation step, the visit count of each ancestor node of $i$ is incremented by one and the cumulative value is also updated to keep consistency.

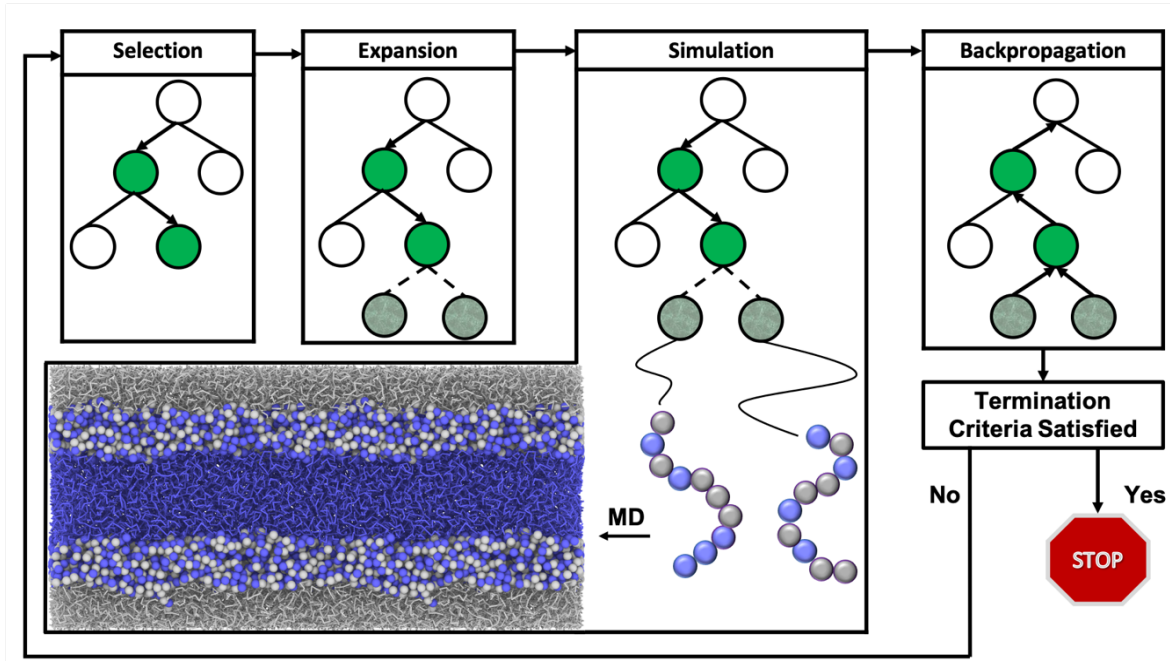

Figure 2: MD-MCTS design scheme for copolymer. It comprise of four steps – selection, expansion simulation and backpropagation that are sequentially conducted as shown by the errors in a given iteration. In the simulation step, MD calculations of a set of candidate structures are conducted parallelly. In the MD snapshot, homopolymers are shown as lines; and copolymers are shown as beads. The termination criteria is chosen to be $\gamma_{12} \approx 0.0$.

**Results:**

We first assess the performance of our MCTS-MD workflow and the same is shown in Figure 4. The lowest surface tension as a function of total number of candidate evaluated during the MCTS iterations is shown in Figure 4a for the three different polymer chain lengths. We find that the MD-MCTS is able to identify optimal sequences that bring down the surface tension to near zero for each of the three cases. An optimal sequence for $\gamma_{12} \approx 0.0$ for all three cases are shown in Figure 4b. The MCTS algorithm is able to achieve zero interfacial energy irrespective of the size or chain length of the polymer compatibilizer. It is interesting to note that all the optimal sequences are non-periodic and non-intuitive.

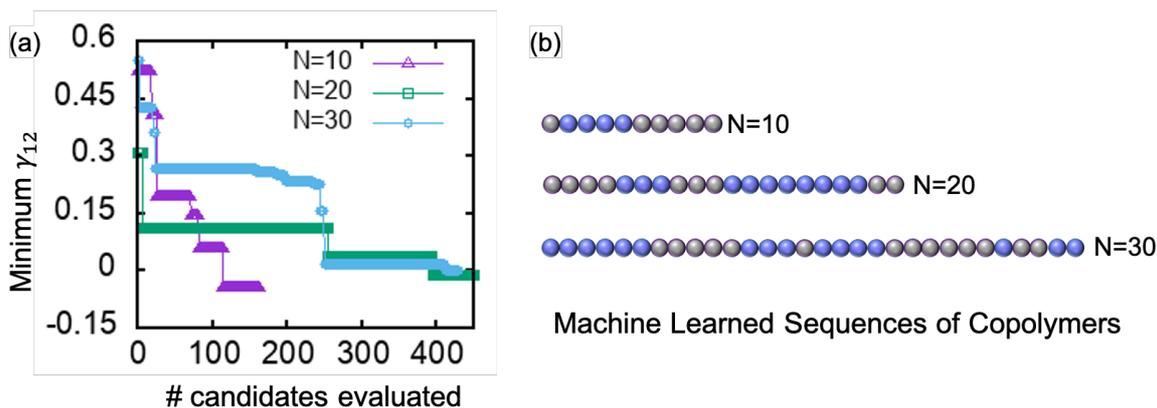

Figure 4: Performance and prediction of MD-MCTS. (a) The lowest surface tension achieved during the MD-MCTS run is shown as a function of total number of candidate materials directly evaluated via MD simulations for all three cases. The optimal sequences for copolymers for all three cases are depicted in (b).

We next test the scalability of the MD-MCTS algorithm by plotting the total number of candidate structures evaluated during a given search cycle as a function of the size of the compatibilizer (Figure 5a). This is especially important considering the long timescale MD simulations that are necessary to perform each evaluation for a specific sequence. MCTS performs direct evaluation of materials properties of 114, 388 and 415 candidates to achieve an optimal sequence that correspond to our target *i.e.* zero interfacial energy of the system for copolymers of length 10, 20 and 30, respectively. We only note a marginal increase in the number of MD simulations with an increase in the system size. This is incredible considering that the design space i.e. the total number of candidates, C increase from 1024 to 1 billion when the chain lengths increase from 10 to 30. MCTS is thus able to attain an optimal solution by screening lower percentage of candidate viz., 22%, 0.07 % and $7.7 \times 10^{-5}$ % of total possible structures for chain length 10, 20 and 30, respectively. Figure 5b depicts this ratio *Cs/C* as a function of *N*, clearly indicating the exponentially lower fraction of candidates required to be screened during the design cycle as polymer chain length increases. This strongly suggests that the MD-MCTS design scheme is scalable to extremely large system sizes, which has hitherto posed a challenge to evolutionary search strategies.

Next, we closely analyse the correlation between any given arrangement and the relative statistics of type-1 and type-0 moieties present in a copolymer chain and the computed surface tension of the system. Mean block length of a copolymer has long been perceived as an important descriptor of a copolymer's properties. Here, we plot the surface tension as a function of mean block length in Figure 6 for each of the three systems (N=10, 20 and 30) simulated during MD-MCTS runs. Here, the mean block length is calculated as the arithmetic mean of the size of all the blocks of 0's and 1's presence in a copolymer. The MCTS is able to identify multiple mean block lengths that correspond to highest performance compatibilizer ($\gamma_{12} \approx 0$). For example, there are copolymers of size N=10 with mean block length $b_l = 1.8$, 2.6 and 3.3 that yield $\gamma_{12} \approx 0$ as shown in Figure 6a. Similar observation can be made for N=20 and N=30 from Figure 6b and 6c, respectively. Moreover, it suggests that sequences with the same mean block length as the optimal sequence can exhibit a wide range of interfacial energy. The interfacial energy varies from 0 to $1.2\varepsilon/\sigma^2$, approximately, for a mean block length of 3.3 for N=10. A similar variation of surface tension for a given mean block is observed for N=20 and 30 as evident in Figure 6b, and 6c, respectively. Thus, the mean block length alone is a poor predictor of a compatibilizer performance as the surface tension varies significantly for copolymers with same mean block length irrespective of polymer size. However, Figure 6 clearly indicates that it is possible to design sequence of two dissimilar moieties, here type-1 and type-0, that will reduce the interfacial energy to zero of a polymer blend independent of compatibilizer chain length.

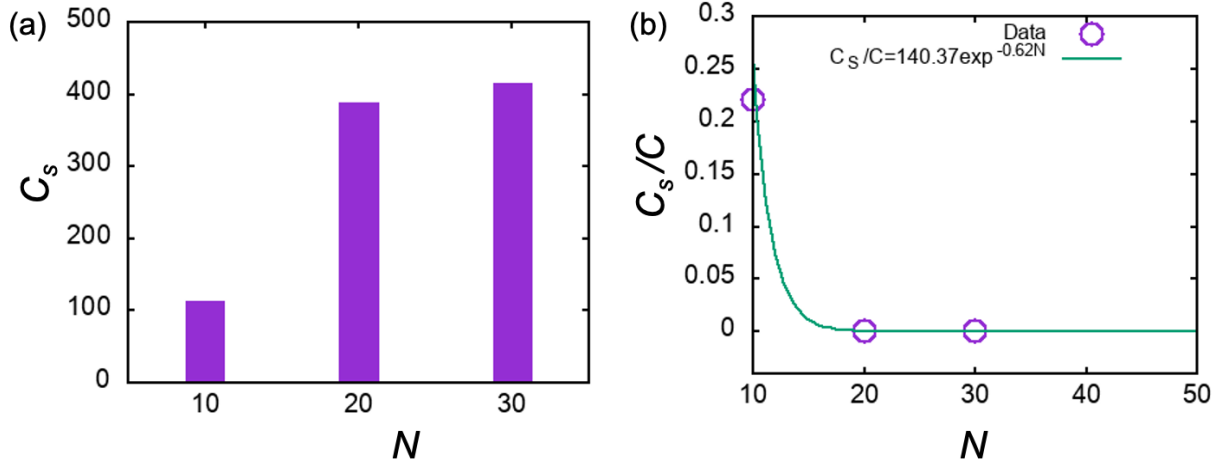

*Figure 5: MD-MCTS complexity. (a) The total number of candidate structures screened($C_S$) during a MCTS cycle is shown as a function of number of monomer in a copolymer chain. (b) The fraction of candidate structures ($C_S/C$) screened during a design cycle is plotted as a function of polymer chain length. for The coefficient of determinant of the fitted lines are $R^2=0.82$ and 0.99 for (a), and (b), respectively.*

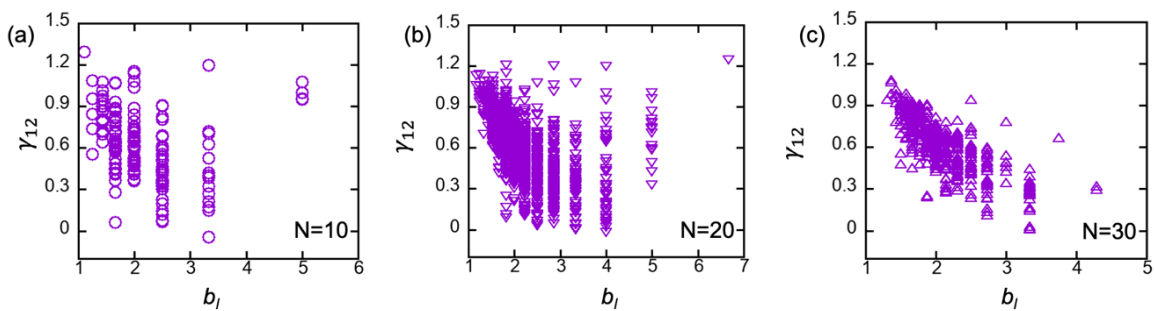

*Figure 6: Variation of interfacial energy is shown as a function of the mean block length for all the candidate structures screened in this study for N = 10, 20 and 30 in (a), (b) and (c), respectively. Both the surface tension and mean block length are in LJ unit.*

In order to further understand the uniqueness of the optimal polymer sequence, we study the interrelationship between mean block length, monomer mole fraction and the interfacial energy of the system. Here, the mole fraction is defined as the ratio of the number of monomer of one particular moiety viz. type-1 moiety and the total number of monomers in a chain. Figure 7a, b and c show the variation of interfacial energy for N=10, 20 and 30, respectively, as a function of mean block length ($l_b$) and mole fraction of type-1 (*q*). The deep blue region in Figure 7 corresponds to lowest surface tension of the system. For N=10, we observe $\gamma_{12} \approx 0.0$ for $q \approx 0.5$ and $l_b = 3.3$. As N increases, there are larger patches of isolated blue regions that appear in the $q$-$l_b$ surface, indicating greater number of global optimal points in the $q$-$l_b$ surface. As the chain length further increases, it is possible to achieve many more sequences that can lead to zero interfacial energy of the system. Often times, the optimal regions are separated by regions of suboptimal points as evident from Figure 7b and 7c. Many of the local optimization strategies and even global ones struggle to navigate around these sub-optimal points. Thus the complexity as well as the degeneracy of this configurational space increases for longer chain length and poses challenges. The MCTS algorithm is however able to effectively navigate around these sub-optimal regions by growing other branches of the tree effectively utilizing the trade-off mechanism between exploration and exploitation. MCTS thus simultaneously explores potentially better pathways to reach the optimal point in a search space and exploits pathways that have the greatest estimate value of the search function. This combination of exploration *vs.* exploitation and an appropriate trade-off mechanism between them, represents a powerful strategy of identifying optimal point for a given function.

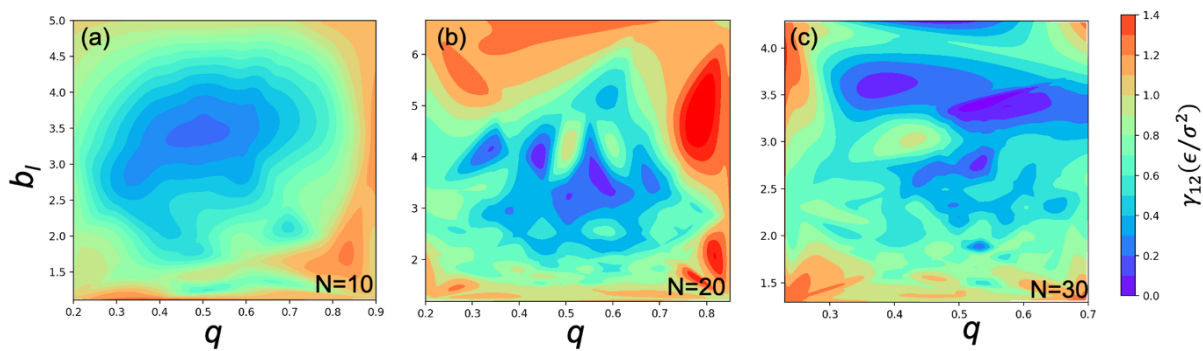

*Figure 7: Heat map of interfacial energy. The interfacial energy $\gamma_{12}$ is shown as a function of mole fraction of A type monomer, q, and mean block length, $l_b$. The surface tension and block length are in LJ unit.*

**Conclusions**

In summary, sequence control in soft matter systems such as polymers and proteins has been a longstanding goal and is highly desirable for a wide variety of applications. In particular, the emergence of sequence control polymers provides tremendous opportunities for material design. By controlling the sequence of a copolymer, one can reliably tune their functionality over a wide range which can be mapped on to their chemical information. The vast combinatorial search space that is required to be explored for these sequence problems pose a major challenge. To overcome this, we interface a gaming AI algorithm viz., MCTS with an MD simulator to tackle the complexity of copolymer design. Our MD-MCTS design algorithm is employed to identify optimal copolymers sequences which can be used as compatibilizer to improve thermodynamics stability of two immiscible homopolymer blends. Unlike other molecular inverse design strategies *i.e.* genetic algorithm (GA) or Bayesian optimization (BO) where design time grows rapidly with the system size, MCTS interfaced with MD appears to require relatively much smaller number of candidate evaluations in any given design cycle. We show that one can engineer the sequence of chemical moieties that will nullify the interfacial tension between immiscible polymers irrespective of the size of compatibilizer polymer chain. Our work also elucidates the correlation between interfacial energy and the sequence statistics of copolymer compatibilizer molecules and illustrates the complexities associated with sequence control over larger polymer chain lengths. Finally, we show that MCTS is highly scalable and efficiently able to navigate large design spaces typical of most practical sequence control related design problem. More broadly, the work provides new strategy that can be used for sequence control and inverse design of copolymers for materials applications.

**Acknowledgement**

The use of the Center for Nanoscale Materials, an Office of Science user facility, was supported by the U.S. Department of Energy, Office of Science, Office of Basic Energy Sciences, under Contract No. DE-AC02- 06CH11357. This research used resources of the National Energy Research Scientific Computing Center, which was supported by the Office of Science of the U.S. Department of Energy under Contract No. DE-AC02-05CH11231. An award of computer time was provided by the Innovative and Novel Computational Impact on Theory and Experiment (INCITE) program of the Argonne Leadership Computing Facility at the Argonne National Laboratory, which was supported by the Office of Science of the U.S. Department of Energy under Contract No. DE-AC02-06CH11357. The authors would like to acknowledge the support from the Argonne LDRD-2017-012-N0 project, UIC faculty start-up fund and IIT Madras faculty stat-up fund.